\documentclass{PoS}

\usepackage{graphicx}
\usepackage{booktabs}
\usepackage{rotating}
\usepackage{listings}
\usepackage{amsmath}
\usepackage{amsfonts}
\usepackage{amssymb}
\usepackage{color}
\usepackage{multirow}

\title{Excited spectroscopy of mesons containing charm quarks from lattice QCD}

\ShortTitle{Excited spectroscopy of mesons containing charm quarks from lattice QCD}

\author{\speaker{Graham Moir} $^{a,b}$ ~ Michael Peardon$^a$ ~ Sin\'{e}ad M. Ryan$^a$ ~ Christopher E. Thomas$^{a,c}$ ~ Liuming Liu$^d$ \\

        $^a$ School of Mathematics, Trinity College, Dublin 2, Ireland \\
	$^b$ Department of Physics, Bergische Universit\"{a}t Wuppertal, Gaussstr. 20, D-42119 Wuppertal, Germany \\
	$^c$ DAMTP, University of Cambridge, Centre for Mathematical Sciences, Wilberforce Road, Cambridge CB3 0WA, UK \\	
        $^d$ Helmholz-Institut f\"{u}r Strahlen-und Kernphysik and Bethe Center for Theoretical Physics, Universit\"{a}t Bonn, D-53115 Bonn, Germany \\

E-mail: \email{moirg@tcd.ie}
\\
\\
(For the Hadron Spectrum Collaboration)
}

\abstract{We present highly excited spectra of charmonium, charm-light and charm-strange mesons using the dynamical anisotropic lattices of the Hadron Spectrum Collaboration. The use of novel techniques has allowed us to extract these spectra with a high degree of statistical precision, while also enabling us to observe states as high as spin-$4$, states with exotic quantum numbers and candidate gluonic excitations. We discuss the phenomenology of these spectra, and also present preliminary results of $D\pi$ scattering considering only the lowest partial wave in the isospin-$3/2$ channel.}

\FullConference{31st International Symposium on Lattice Field Theory, LATTICE 2013 \\
                 July 29 - August 3, 2013 \\
                 Mainz, Germany}

\begin{document}

\section{Introduction}

The discovery of a plethora of narrow charmonium-like resonances, commonly known as the `X, Y, Z' states, and the observation of the enigmatic $D^{*}_{s0}(2317)^{\pm}$ and $D_{s1}(2460)^{\pm}$ states, has forced theorists to re-examine the once `understood' charm sector. While the nature of these states remains hitherto unexplained, many interesting classifications have been put forward. These include molecular mesons, hybrids (a $q\bar{q}$ pair with a gluonic excitation) and tetraquarks.

In these proceedings, we give an overview of our previously published results in the charmonium, charm-light and charm-strange sectors \cite{Liu, Moir}. We also present preliminary results of the isospin-$3/2$, $D\pi$ elastic scattering phase shift. Our calculations are performed on dynamical $N_{f} = 2 + 1$ anisotropic ensembles generated by the Hadron Spectrum Collaboration \cite{Edwards, Lin}. The light quarks are unphysically heavy corresponding to a pion mass of $M_{\pi} \approx 391$ MeV. The anisotropy is such that the spatial and temporal lattice spacings are related via $\xi = a_{s}/a_{t} \sim 3.5$, ensuring that we simulate with $a_{t}m_{c} \ll 1$ and that the standard relativistic formulation of fermions can be used for the charm quark. For a more detailed discussion see \cite{Liu}.

Table \ref{table:charm_ensembles} summarises the ensembles used in our calculation of the charmonium spectrum, Table \ref{table:open_charm_ensembles} summarises the ensembles used in our calculation of the charm-light and charm-strange spectra, while Table \ref{table:dpi_ensembles} summarises the ensembles used in the calculation of the isospin-$3/2$, $D\pi$ elastic scattering phase shift. 

\begin{table}[b]
\begin{center}
\begin{tabular}{|ccccccc|}
\hline
Volume & & $N_{cfgs}$ & & $N_{tsrcs}$ & & $N_{vecs}$ \\
\hline
$16^{3} \times 128$ & & 96 & & 128 & & 64 \\
$24^{3} \times 128$ & & 552 & & 32  & & 162 \\
\hline
\end{tabular}
\caption{The gauge field ensembles used in the calculation of the charmonium spectrum. $N_{cfgs}$ and $N_{tsrcs}$ are respectively the number of gauge field configurations and time-sources per configuration used; $N_{vecs}$ refers to the number of eigenvectors used in the distillation method \cite{Peardon}.}
\label{table:charm_ensembles}
\end{center}
\end{table}

\begin{table}[t]
\begin{center}
\begin{tabular}{|ccccccc|}
\hline
Volume & & $N_{cfgs}$ & & $N_{tsrcs}$ & & $N_{vecs}$ \\
\hline
$16^{3} \times 128$ & & 96 & & 128 & & 64 \\
$24^{3} \times 128$ & & 553 & & 16 & & 162 \\
\hline
\end{tabular}
\caption{As for Table 1, but for the calculation of the charm-light and charm-strange spectra.}
\label{table:open_charm_ensembles}
\end{center}
\end{table}

\begin{table}[b]
\begin{center}
\begin{tabular}{|ccccccc|}
\hline
Volume & & $N_{cfgs}$ & & $N_{tsrcs}$ & & $N_{vecs}$ \\ 
\hline
$20^{3} \times 128$ & & 603 & & 3 & & 128 \\
\hline
\end{tabular}
\caption{As for Table 1, but for the calculation of the $D\pi$ phase shift.}
\label{table:dpi_ensembles}
\end{center}
\end{table}

\section{Spectroscopic Extraction}\label{sec:spectroscopy}

In order to maximise the spectral information we can extract from two-point correlation functions, we employ a large basis of \textit{distilled} \cite{Peardon} fermion bilinear interpolating operators; we use the same derivative based construction described in \cite{Dudek}, including operators with up to three derivatives. This allows access to all combinations of $J^{P(C)}$ with $J \leq 4$ (except for the exotic $4^{+-}$ combination); here $J$ is the total angular momentum, while $P$ and $C$ are the parity and charge conjugation quantum numbers respectively. It should be noted that $C$ is only a good quantum number for flavour singlets, i.e. it is not a good quantum number for charm-light and charm-strange combinations. 

For each lattice irrep, $\Lambda^{P(C)}$, and flavour sector, we compute an $N \times N$ matrix of correlation functions,
\begin{equation}\label{corr}
C_{ij}(t) \equiv \langle 0|{\cal O}_i(t){\cal O}^{\dagger}_j(0)|0\rangle = \sum_\mathfrak{n} \frac{Z_i^{\mathfrak{n}*} Z_j^{\mathfrak{n}}}{2E_\mathfrak{n}} e^{-E_\mathfrak{n}t} ~, 
\end{equation}
where $N$ is the number of operators in a given lattice irreducible representation (irrep). The sum is over all states with the same quantum numbers as the operators ${\cal O}^{\dagger}(0)$ and ${\cal O}(t)$, and the operator-state overlaps $Z^{\mathfrak{n}}_i \equiv \langle \mathfrak{n} | \mathcal{O}^{\dagger}_i | 0 \rangle$. We extract energies via the use of a variational technique, which practically amounts to solving
\begin{equation}
C_{ij}(t) v^{\mathfrak{n}}_j = \lambda^{\mathfrak{n}}(t,t_0) C_{ij}(t_0) v^{\mathfrak{n}}_{j} ~,
\end{equation}
where an appropriate reference time-slice, $t_{0}$, must be chosen as described in \cite{Dudek}. The energies are obtained by fitting the dependence of the eigenvalues, $\lambda^{\mathfrak{n}}$, on $(t - t_{0})$. The eigenvectors $v^{\mathfrak{n}}$ are related to the operator-state overlaps, $Z^{\mathfrak{n}}$, and play a vital role in the spin identification of states as described in \cite{Dudek}.

\section{Hidden and Open-Charm Spectra}\label{sec:spectra}     

A simple non-relativistic picture of a meson describes it as a $q\bar{q}$ bound state whose spins are coupled together to form a total spin $S$, which is then coupled to an orbital angular momentum $L$ in order to create a state of total angular momentum $J$. If the meson is an eigenstate of charge-conjugation, some $J^{PC}$ combinations are inaccessible in this description; such states are termed \textit{exotic}. The supermultiplet structure expected from this model is $n^{2S + 1}L_{J}$, where $n$ is the radial quantum number.

\begin{figure}[t]
\begin{center}
\includegraphics[width=0.85\textwidth]{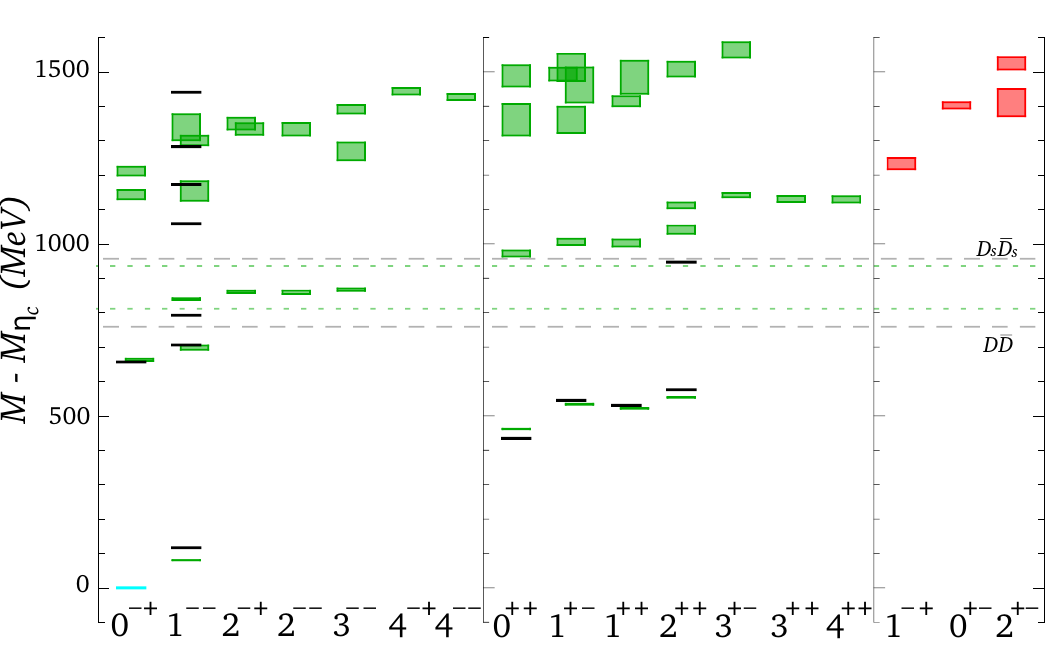}
\caption{Our calculated charmonium spectrum labelled by $J^{PC}$. The red and green boxes are the masses calculated on the $24^{3}$ volume; black lines are experimental values from the PDG \cite{PDG}. We show the calculated (experimental) masses with the calculated (experimental) $\eta_{c}$ mass subtracted. The vertical size of the boxes represents the one-sigma statistical uncertainty. The dashed lines indicate the lowest non-interacting $D\bar{D}$ and $D_{s}\bar{D}_{s}$ levels using the $D$ and $D_{s}$ masses calculated on the $16^{3}$ volume (fine green dashing) and using the experimental masses (coarse grey dashing).}
\label{fig:charmonium_final} 
\end{center}
\end{figure}

Figure \ref{fig:charmonium_final} shows our final charmonium spectrum calculated on our $24^{3}$ volume; in general we find no significant volume dependence when comparing to our $16^{3}$ volume. Most of the states with non-exotic $J^{PC}$ seem to follow the simple $n^{2S + 1}L_{J}$ pattern; we assign states to a particular supermultiplet based on their operator-state overlap values as described in \cite{Liu, Dudek1}. In the left panel of Figure \ref{fig:charmonium_final} we show negative parity states with non-exotic $J^{PC}$. Of these, there are three states with $J^{PC} = (0, 2)^{-+}, 1^{--}$ which do not appear to fit into the $n^{2S + 1}L_{J}$ supermultiplet structure. Moreover, we find that these states are almost degenerate with the lightest exotic meson, which has $J^{PC} = 1^{-+}$; states with exotic $J^{PC}$ are shown in the rightmost panel of Figure \ref{fig:charmonium_final}. Our extracted overlap values for these four states indicate that they have a common structure \cite{Liu, Dudek1}, and on this basis we suggest that they form the lightest hybrid supermultiplet in the charmonium sector. 

The middle panel of Figure \ref{fig:charmonium_final} shows the non-exotic positive parity states. High up in the spectrum we observe eight states that have a large overlap onto operators proportional to the field strength tensor, and so we interpret them as non-exotic hybrid mesons. There are three positive parity exotic states similar in mass with $J^{PC} = 0^{+-}, 2^{+-}$. These also have large overlap onto operators proportional to the field strength tensor. By considering their operator-state overlaps, we interpret these states as members of the first excited hybrid supermultiplet in the charmonium sector. Interestingly, the two observed hybrid supermultiplets fit the pattern expected by a $1^{+-}$ gluonic excitation coupled to a quark-antiquark pair, such as in the $P$-wave quasi-gluon approach. For more details see \cite{Liu, Dudek1}.

\begin{figure}[t]
\begin{center}
\includegraphics[width=0.9\textwidth]{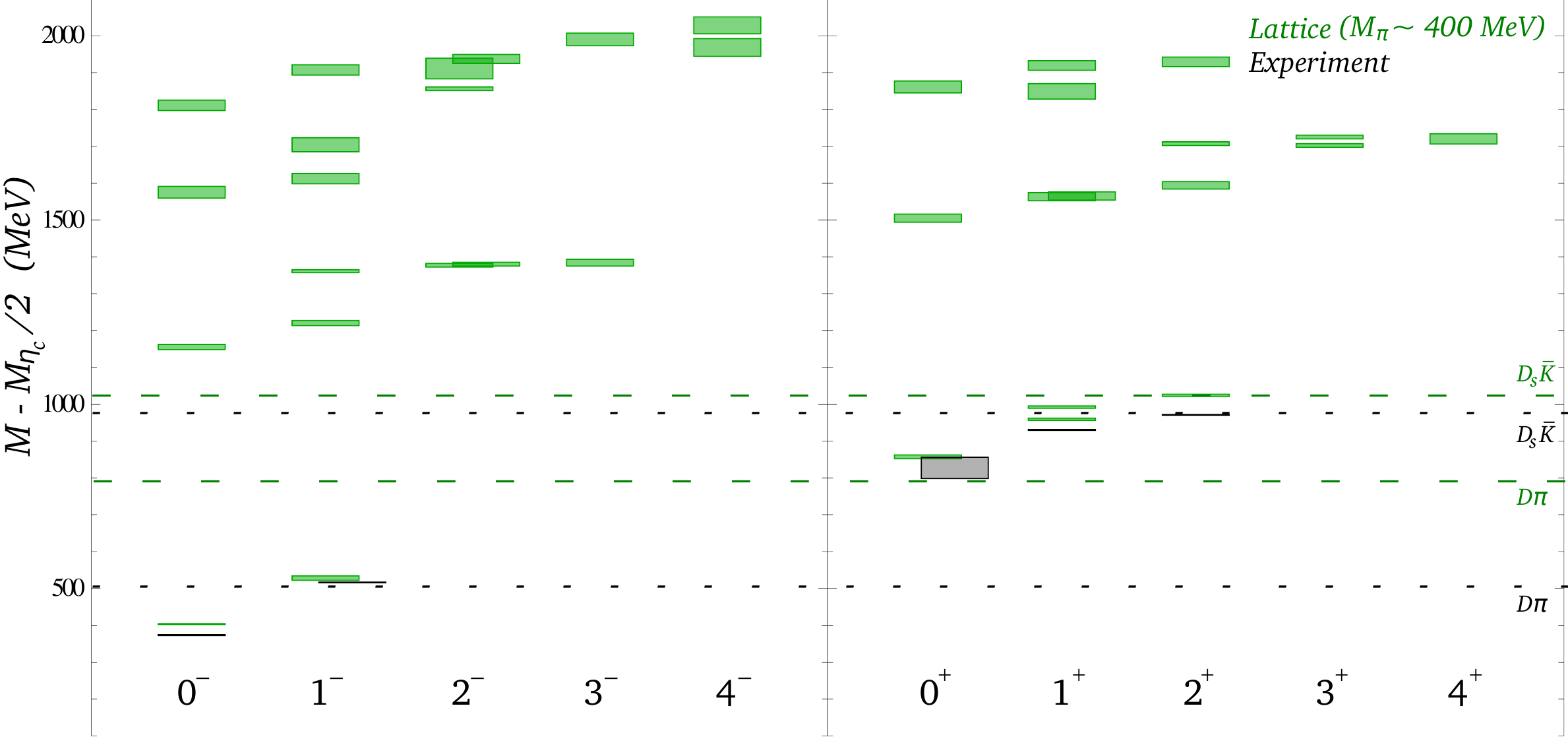}
\caption{Our calculated charm-light spectrum labelled by $J^{P}$. The green boxes are our calculated masses, while the black boxes correspond to experimental masses of neutral charm-light mesons \cite{PDG}. We present the calculated (experimental) masses with half the calculated (experimental) $\eta_{c}$ mass subtracted to reduce the uncertainty from tuning the bare charm-quark mass. The vertical size of each box indicates the one sigma statistical uncertainty on either side of the mean. The dashed lines show the lowest non-interacting $D\pi$ and $D_{s}\bar{K}$ thresholds using our measured masses (coarse green dashing) and experimental masses (fine black dashing).}
\label{fig:D_final}
\end{center}
\end{figure}

\begin{figure}[t]
\begin{center}
\includegraphics[width=0.9\textwidth]{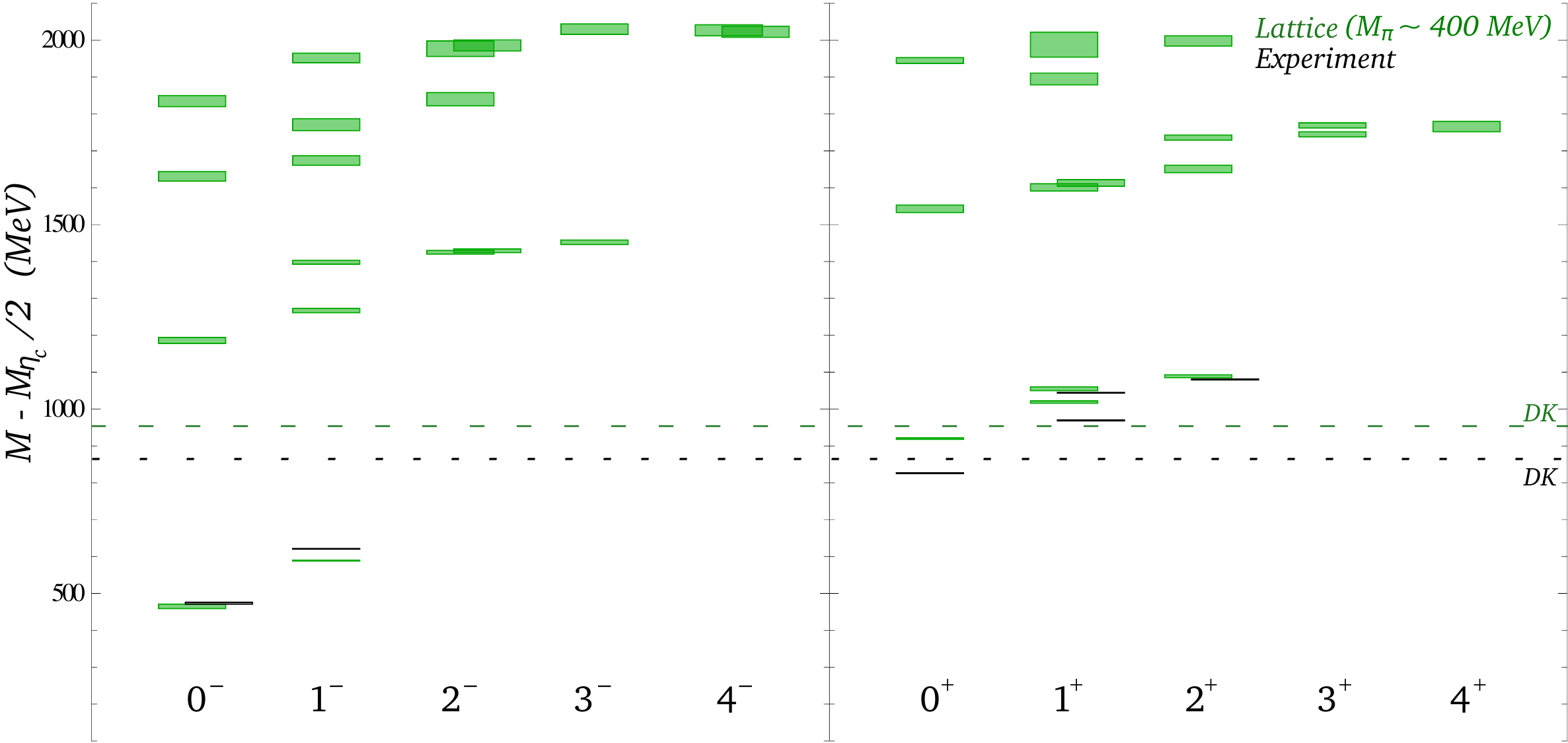}
\caption{As for Figure 2, but for the charm-strange combination. Here the thresholds correspond to the lowest non-interacting $DK$ energy.}
\label{fig:Ds_final}
\end{center}
\end{figure}

Figures \ref{fig:D_final} and \ref{fig:Ds_final} show our final charm-light and charm-strange spectra respectively as calculated on our $24^{3}$ volume. In general, throughout both spectra we find no significant volume dependence when comparing with our $16^{3}$ results. However, there are a couple of exceptions: for the lightest $0^{+}$ and $1^{+}$, determined with high statistical precision, we find a $2\sigma$ discrepancy between the two volumes. These states lie precariously close to thresholds and therefore mixing with multi-meson states may be important and could be the cause of the observed volume dependence. In these calculations, our operator bases do not include any operators that `look like' two-mesons and so, we do not expect to be able to reliably extract multi-meson energy levels; a conservative approach is to suggest that our mass values are accurate only up to the hadronic width \cite{Moir}.

As in the charmonium case, most of our calculated states fit into a $n^{2S + 1}L_{J}$ pattern. Around $1.2$ GeV above the ground state in the negative parity sector of both spectra, we observe four states that do not appear to fit into the $n^{2S + 1}L_{J}$ pattern. Due to their relatively strong overlap onto operators proportional to the field strength tensor, we interpret these states as hybrid mesons. As in the charmonium sector, we suggest that these four states form the lightest hybrid supermultiplets in their respective sectors due to their similar operator-state overlap values \cite{Moir}.

\section{$D\pi$ Elastic Scattering Phase Shift in the Isospin-$3/2$ Channel}

It is clear from Figures \ref{fig:D_final} and \ref{fig:Ds_final} that multi-hadron effects may play an important role in both spectra; the lightest $P$-wave supermultiplet lies around the lowest non-interacting thresholds, which are the $D\pi$ and $D_{s}K$ thresholds in the charm-light sector and the $DK$ threshold in the charm-strange sector. As previously mentioned in section \ref{sec:spectra}, we find a $2\sigma$ discrepancy between our $16^{3}$ and $24^{3}$ determination of the lightest $0^{+}$ and $1^{+}$ states in both sectors, further motivating our study of multi-hadron and scattering states.  

We perform our calculation of the $D\pi$ elastic scattering phase shift in the isospin-$3/2$ channel, on our $20^{3}$ volume with the parameters described in Table \ref{table:dpi_ensembles}. In order to calculate the discrete finite volume $D\pi$ spectrum, we follow Ref. \cite{Dudek2} and construct a large basis of \textit{optimised operators} that `look like' two-mesons and that overlap predominantly with a $D\pi$ combination. These optimised operators are themselves built from linear combinations of the operators discussed in section \ref{sec:spectroscopy} and the so-called \textit{helicity operators} of Ref. \cite{Thomas}. In our calculation, we include optimised $D\pi$ combinations up to an overall momentum of $P^{2} = 3$ corresponding to a variety of internal momenta.

\begin{figure}[t]   
\begin{center}
\includegraphics[width=0.8\textwidth]{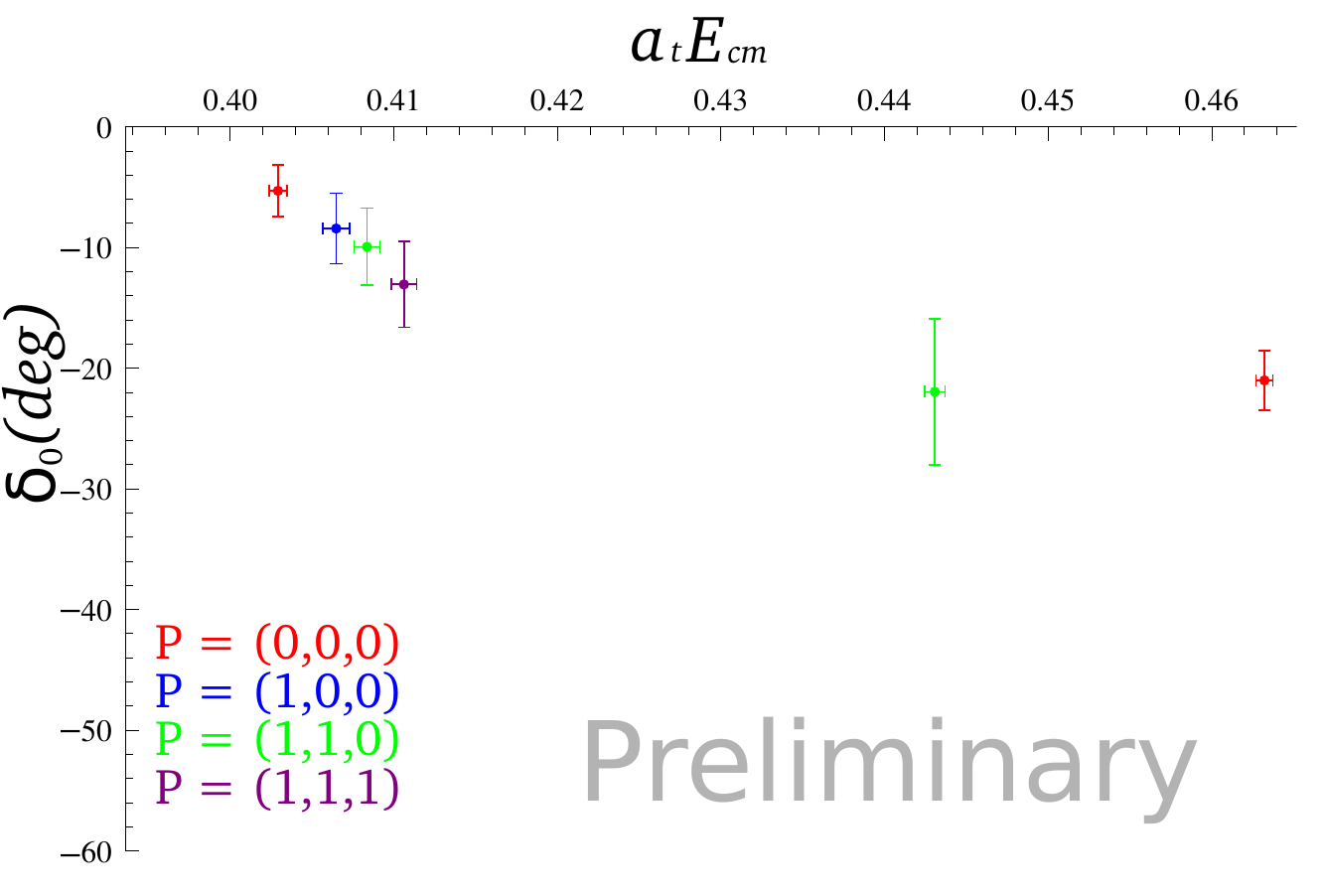}
\caption{The phase shift, $\delta_{0}$, in the elastic scattering region determined from the $D\pi$ spectrum calculated on our $20^{3} \times 128$ volume. The error bars represent the statistical uncertainty. The colour coding indicates the overall momentum, $\vec{P}$, of the state from which a given data point was extracted. The centre-of-momentum frame energies are given in units of $2\pi/L_{s}$, where $L_{s}$ is the spatial extent of the lattice. The inelastic threshold lies at $a_{t}E_{cm} \approx 0.47$.}
\label{fig:phase_shift}
\end{center}
\end{figure}

To compute the phase shift, we apply the L\"{u}scher formalism \cite{Luscher} and its extension to moving frames \cite{Rummukainen} to the $D\pi$ multi-meson spectra. For each overall momentum, $\vec{P}$, we only consider the lowest partial wave, $l = 0$. However, at modest momenta $\delta_{0} \ll \delta_{2} \ll \delta_{4} ...$, so we do not expect the effects of higher partial waves to dramatically change the overall qualitative picture. Figure \ref{fig:phase_shift} shows the $D\pi$ elastic scattering phase shift, $\delta_{0}$, as a function of the centre-of-momentum frame energies, which are given in units of $2\pi/L_{s}$, where $L_{s}$ is the spatial extent of the lattice. Resonances show up in the phase shift as a rapid variation through $180$ degrees. It is clear from Figure \ref{fig:phase_shift} that we see no such behaviour, and hence no resonance. The shape of our phase shift is consistent with that of a weakly repulsive interaction, which is to be expected in the $I = 3/2$ channel.

Our study of $D\pi$ scattering is ongoing and the obvious next step is to increase the number of points used to map out the phase shift shown in Figure \ref{fig:phase_shift} by extending our study to include our $16^{3}$ and $24^{3}$ volumes. While we also plan to study the $D\pi$ isospin-$1/2$ channel, calculations of $DK$ scattering are already under way and may help to shed some light on the nature of the enigmatic $D_{s0}^{*}(2317)^{\pm}$ and $D_{s1}^{*}(2460)^{\pm}$ states.

\section*{Acknowledgements}
We thank our colleagues within the Hadron Spectrum Collaboration. {\tt Chroma} \cite{Edwards1} and {\tt QUDA} \cite{Clark, Babich} were used to perform this work on the Lonsdale cluster maintained by the Trinity Centre for High Performance Computing funded through grants from Science Foundation Ireland (SFI), at the SFI/HEA Irish Centre for High-End Computing (ICHEC), and at Jefferson Laboratory under the USQCD Initiative and the LQCD ARRA project. Gauge configurations were generated using resources awarded from the U.S. Department of Energy INCITE program at the Oak Ridge Leadership Computing Facility at Oak Ridge National Laboratory, the NSF Teragrid at the Texas Advanced Computer Center and the Pittsburgh Supercomputer Center, as well as at Jefferson Lab. This research was supported by the European Union under Grant Agreement number 238353 (ITN STRONGnet) and by the Science Foundation Ireland under Grant Nos.~RFP-PHY-3201 and RFP-PHY-3218. GM acknowledges support from the School of Mathematics at Trinity College Dublin. CET acknowledges support from a Marie Curie International Incoming Fellowship, PIIF-GA-2010-273320, within the 7th European Community Framework Programme.

\end{document}